\documentstyle[11pt]{article}
\begin{document}
\title{Gauge Theories in a Bag}
\author{A. Wipf\\
Institute for Theoretical Physics, ETH-H\"onggerberg,\\
CH-8093 Z\"urich, Switzerland\\
\\
S. Duerr\\
Institute for Theoretical Physics, University of Z\"urich,
Winter-\\thurerstr. 190, CH-8057 Z\"urich, Switzerland}
\date{ZU-TH 30/94 and ETH-TH/94-36\\
15.11.1994}
\maketitle

\newcommand{\eqnn}[1]{\begin{eqnarray*}#1\end{eqnarray*}}
\newcommand{\eqnl}[2]{\par\parbox{11cm}
{\begin{eqnarray*}#1\end{eqnarray*}}\hfill
\parbox{1cm}{\begin{eqnarray}\label{#2}\end{eqnarray}}}
\newcommand{\eqnlb}[2]{\begin{equation}\fbox{$\displaystyle
#1 $}\label{#2}\end{equation}}
\newcommand{\eqngr}[2]{\par\parbox{11cm}
{\begin{eqnarray*}#1\\#2\end{eqnarray*}}\hfill
\parbox{1cm}{\begin{eqnarray}\end{eqnarray}}}
\newcommand{\eqngrlb}[3]{\par\parbox{11cm}
{\begin{eqnarray}\fbox{$\displaystyle   #1\\#2$}\end{eqnarray}}\hfill
\parbox{1cm}{\begin{eqnarray}\label{#3}\end{\eqnarray}}}

\newcommand{\eqngrl}[3]{\par\parbox{11cm}
{\begin{eqnarray*}#1\\#2\end{eqnarray*}}\hfill
\parbox{1cm}{\begin{eqnarray}\label{#3}\end{eqnarray}}}

\newcommand{\eqngrn}[3]{\begin{eqnarray*}#1 \\ #2\end{eqnarray*}}

\newcommand{\eqngrr}[3]{\par\parbox{11cm}
{\begin{eqnarray*}#1\\#2\\#3\end{eqnarray*}}\hfill
\parbox{1cm}{\begin{eqnarray}\end{eqnarray}}}

\newcommand{\eqngrrl}[4]{\par\parbox{11cm}
{\begin{eqnarray*}#1\\#2\\#3\end{eqnarray*}}\hfill
\parbox{1cm}{\begin{eqnarray}\label{#4}\end{eqnarray}}}

\newcommand{\refs}[1]{(\ref{#1})}

\def\Box{\Delta}
\def\tsi{\tilde\sigma}
\def\hb{\hbar}
\def\dett{{\det}_\theta}
\def\di{D\!\!\!\!\slash\,}
\def\sa{a\!\!\!\!\slash\,}
\def\fdi{\partial\!\!\!\slash\,}
\def\ddx{d^dx}
\def\tr{\,{\rm tr}\,}
\def\bet{\bar\eta}
\def\bal{\bar\alpha}
\def\bps{\bar\psi}
\def\gan{\gamma_n}
\def\cc{{\cal C}}
\def\cd{{\cal D}}
\def\cs{{\cal S}}
\def\>{\rangle}
\def\<{\langle}
\def\gf{\gamma_5}
\def\gff{\bar\gamma}
\def\pan{\par\noindent}
\def\cov{\bigtriangledown}
\def\pd{\psi^{\dagger}}
\def\mtxt#1{\quad\hbox{{#1}}\quad}
\def\pa{\partial}
\def\gam{\gamma}
\def\eps{\epsilon}
\def\es{\!=\!}
\def\lapf{\triangle_f}
\def\lap{\triangle}
\def\olap{{1\ov\lap}}
\def\ov{\over}
\def\cd{{\cal D}}
\def\om{\omega}
\def\pamu{{\partial_\mu}}
\def\panu{{\partial_\nu}}
\def\al{\alpha}
\def\be{\beta}
\def\si{\sigma}
\def\pr{\prime}
\def\lam{\lambda}
\def\tr{\hbox{tr}\,}
\def\cm{{\cal M}}
\def\cf{{\cal F}}
\def\cmb{{\pa{\cal M}}}
\def\ha{{1\over 2}}
\begin{abstract}
We investigate multi-flavour gauge theories confined in $d\es 2n$-dimensional
Euclidean bags. The boundary conditions for the 'quarks'
break the axial flavour symmetry and depend on a parameter $\theta$.
We determine the $\theta$-dependence of the fermionic correlators
and determinants and find that a $CP$-breaking $\theta$-term is generated
dynamically. As an application we calculate the chiral
condensate in multi-flavour $QED_2$ and the abelian projection
of $QCD_2$. In the second model a condensate is generated
in the limit where the number of colours, $N_c$, tends to infinity.
We prove that the condensate in $QCD_2$ decreases with increasing
bag radius $R$ at least as $\sim R^{-1/N_cN_f}$.
Finally we determine the correlators of mesonic currents in $QCD_2$.
\end{abstract}
\section{Introduction}
Possible mechanisms for the spontaneous breaking of the chiral symmetry
in $QCD$ have repeatedly been discussed in the literature \cite{shuryak},
but a derivation from first principles
remains to be found. The broken phases can be probed
by coupling the fields to a symmetry breaking trigger source
which is removed after the infinite volume limit has been taken.
Alternatively one may put the system in a finite box,
imposes symmetry breaking boundary conditions and
then performs the thermodynamic limit $V\to\infty$.
This is wellknown from spin models \cite{spin}. For example,
when coupling the Ising spins to a constant magnetic field
a mean magnetization remains at low temperature
even when the trigger has been switched off.
Such a magnetization can only arise if the ground state is $Z_2$-asymmetric
or in other words if the $Z_2$-symmetry is spontaneously broken.
Instead of switching on a magnetic field one
may impose $Z_2$-breaking, say spin-up, boundary conditions and again
a magnetization remains after the infinite
volume limit has been taken.\par
In $QCD$ a great deal of efforts have been undertaken to study the
quark condensates in the limit of vanishing current quark masses
\cite{shuryak}. These condensates would signal a spontaneous breaking of the
axial flavour symmetry $SU_A(N_f)$ as it is required by the
low energy phenomenology.
Here one runs into the following paradox: In the chiral limit
the generating functional for the fermionic Green's functions
on a compact spacetime without boundary,
\eqngrl{
Z[\eta,\bar\eta]&=&\int\cd (A,\psi)\,e^{-S_{YM}+\int\bar\psi i\di\psi
+\int \bar\eta\psi+\bar\psi\eta}}
{&=&\sum_N\int\cd A^N\, e^{-S_{YM}}\prod_{k=1}^N(\bar\eta,\psi_k)
(\bar\psi_k,\eta){\det}^\pr i\di\;e^{\int \bar\eta S^\pr\eta},}{ein1}
where the gauge fields $A^N$ support $N$ zero modes
$\psi_1,\dots,\psi_N$ of $i\di$, gets contributions from sectors
with non-zero instanton numbers \cite{tH}
\eqnl{
q={1\ov 32 \pi^2}\int d^4x F_{\mu\nu}^a\;^*F_{\mu\nu}^a.}{ein2}
The primes in \refs{ein1} indicate the suppression of zero modes.
If we only allow for smooth configurations on $S^4$ or
$S^3\times R$ then  $q$ is an integer \cite{PolSchw} and the
number of zero modes \cite{Schw}
\eqnl{N=
\left\{ \begin{array}{ll}
N_f q &\hbox{for }N_f\hbox{-flavour }QCD\\
N_c q &\hbox{for supersymmetric }QCD \end{array} \right.   }{ein2a}
is an integer multiple of $N_f$ or $N_c$. Thus neither
the topologically trivial sector contributes to the chiral condensate
\eqnl{
\<\bar\psi \psi\>={1\ov Z}{\delta^2\ov \delta \eta\delta\bar\eta}
Z\vert_{\eta=\bar\eta=0},}{ein2b}
since $S^\pr$ in \refs{ein1} is chirality conserving, nor
the nontrivial sectors since there are too many zero modes.
Hence the condensate vanishes\footnote{When
switching on a small quark mass one arrives
at the same conclusion on a compact spacetime without boundary,
since $\det(i\di+m)\sim m^N$.}. This
conclusion is certainly in conflict with low energy strong interaction
phenomenology or Ward-identities which predict a nonvanishing
condensate for susy $QCD$ \cite{NovShif}.\par
Possible ways out (which work if the center of the gauge group
is big enough) have been suggested by t'Hooft \cite{torons},
who introduced twisted instantons, socalled torons, on the $4$-dimensional
torus, and by Zhitnitsky \cite{Zhitnitsky},
who considered singular gauge fields on $S^4$.
Both constructions produce configurations with fractional
instanton numbers and may resolve the above mentioned paradox.
However, for $O(N\!>\!4)$ susy-YM-theories, which give rise to a nonvanishing
chiral condensate \cite{ShifVain}, the center is too small and
these constructions do not work. Recently Shifman and
Smilga have introduced another type of configuration, they called
them fractons, which may generate a chiral condensate \cite{ShifSmil}.
By allowing for flavour-dependent twisted boundary conditions
they could introduce fractionally charged instantons and those
generated a non-vanishing condensate in multi-flavour $QED_2$.
It remains to be seen whether these fractons solve the puzzle posed
by the chiral condensate in $O(N)$-susy theories. \par
Instead of quantizing gauge theories on a sphere or on a torus
we propose to quantize them in an even-dimensional ($d\es 2n$)
Euclidean bag $\cm$ \cite{bc1} and to impose $SU_A(N_f)$-breaking boundary
conditions to trigger a chiral symmetry breaking.
In a bag the instanton number is not quantized and the
system itself is allowed to decide which are the dominant
configurations. We investigate how the various correlators
depend on the parameter $\theta$ characterizing the boundary conditions
and shall see that in the models we studied the bag boundary conditions
are a substitute for small quark masses and also reproduce
the fracton results. \par
In the chiral limit of massless 'quarks' in the fundamental
representation of $SU(N_c)$ the Euclidean action
\eqngrl{
S[A,\psi]=S_{YM}[A]&+&S_D[A,\psi],\mtxt{where}}
{S_{YM}={1\ov 4 g^2}\int\limits_\cm \tr F_{\mu\nu}F^{\mu\nu}
&,&S_D=\sum\limits_{p=1}^{N_f}\int\limits_\cm\pd_p i\di\psi_p,}{ein4}
is invariant under global $SU_V(N_f)\times SU_A(N_f)$
rotations\footnote{Actually, for $N_c\es 2$ the symmetry
group is $SU(2N_f)$ \cite{SV}.}
of the fermions since the Dirac operator
\eqnl{
\di=\gam^\mu D_\mu=\gam^\mu\big(\pa_\mu-iA_\mu\big)}{ein6}
is the same for all $N_f$ flavours. We shall impose
the following boundary conditions, which relate the different
spin components on the bag boundary,
\eqnl{(B(\theta)\times I_f\times I_c) \psi=\psi
\mtxt{on}\cmb.}{ein7}
They break the $SU_A(N_f)$-symmetry
but are vector-flavour and colour neutral so that the
gauge invariant fermionic determinant is the same for all flavours.
This approach has various advantages.
First, the configuration space of gauge potentials
in a bag is topologically trivial and hence there are no
disconnected instanton sectors. Related to that is the
absence of fermionic zero modes which would complicate the
quantization of gauge theories considerably \cite{sw,Jo,jay}. Second,
the $\theta$-dependence of the fermionic determinant, which
appears in the measure of functional integration over the
gauge field configurations after the fermions have been integrated out,
\eqnl{
\<O\>=\int d\mu_\theta (A)\,\<O\>_A\mtxt{,}
d\mu_\theta (A)={1\ov Z}e^{-S_{YM}[A]}{\det}_\theta
(i\di)\;\cd A}{ein8}
can be calculated explicitly, contrary to its mass dependence.
Here $\<O\>_A$ denotes the expectation value of $O$ in
a fixed background gauge field $A$,
\eqnl{
\<O\>_A={1\ov \det_\theta i\di}\int \cd\pd\cd\psi \;O\,e^{\int
\pd i\di\psi}.}{ein9}
In writing \refs{ein8} we anticipated that in a bag $\di$ possesses no
zero modes and absorbed the gauge fixing factor with corresponding
Fadeev-Popov determinant in $\cd A$.\par
The results of our investigations are presented as follows: In section $2$
we introduce the bag boundary conditions for the 'quarks'.
Some simple consequences for the spectrum of the Dirac operator
are then discussed in section 3. We show that $\di$ possesses
no zero modes, discuss the (modified) parity transformation
and derive a boundary Hellmann-Feynman formula.
In section 4 we determine the $\theta$-depencence of the fermionic
Green's functions in a (spherical) bag and find their explicit forms when the
gauge field is switched off. In the following section we derive
the $\theta$-dependence of the fermionic determinants for
arbitrary $2n$-dimensional bags. We shall prove that
through the interaction of the 'quarks' with the boundary
an effective CP-breaking $\theta$-term is generated. In the remaining
part of the paper we investigate $2$-dimensional gauge theories
in the chiral limit.
We start in section 6 with applying the deformation technique to
evaluate the exact fermionic determinant in a bag. We prove that
for $U(N_c)$-theories the measure of functional integration
$d\mu_\theta (A)$ factorizes into the $U(1)$ and $SU(N_c)$ measures.
Then we gain further insight into the spectrum of these models by calculating
all mesonic current correlators in section 7. For $U(N_c)$ gauge
theories with $N_f$ flavours we find that the spectrum contains
$1$ massive and $N_f^2-1$ massless bosons, similarly as in the
multi-flavour Schwinger model, and that they decouple from
the remaining degrees of freedom.
In the last section we investigate the chiral symmetry breaking
in $2$-dimensional gauge theories. First we derive
the exact form of the chiral condensate for multi-flavour
$QED_2$ in a spherical bag. A comparison with the perturbation
by small 'quark'-masses \cite{Smilga1} shows that the
bag-boundary conditions serve as trigger similarly as small 'quark'-masses
do. However, in a bag we need not worry about instantons, torons or fractons.
Then we derive an upper bound on the chiral condensate in nonabelian
gauge theories as a function of the bag-radius. As a particular application we
prove that for $2$-dimensional $SU(N_c)$ gauge theories with arbitrary
$N_c<\infty$ the condensate vanishes in the thermodynamic limit. Finally
we calculate the condensate in the abelian projected
gauge theories and discuss the large $N_c$-limit. We shall see
that for $1$ flavour and $N_c\to\infty$ a condensate is generated.
In the discussion we show that for multi-flavour $QED_2$, confined
in a {\it spatial} bag and at finite temperature,
the chiral condensate agrees with
that generated by fractons on the torus \cite{ShifSmil}.
In the appendix we derive the boundary-Seeley-deWitt coefficient which
is needed in the main body of the paper.

\section{Bag Boundary Conditions}
For Dirac fermions propagating in an Euclidean bag $i\di$ should be
selfadjoint (or at least normal) for the partition function $Z$
to be real. A necessary condition for selfadjointness is that
\eqnl{
(\chi,i\di\psi)-(i\di\chi,\psi)\equiv\int\limits_\cm\chi^\dagger i\di\psi-
\int\limits_\cm(i\di\chi)^\dagger \psi=i\oint\limits_{\cmb}
\chi^\dagger\gan\psi}{bag1}
vanishes. Here $\gan\es n^\mu\gam_\mu$ is the projection
of the hermitean $\gamma$-matrices on the outward oriented normal
vectorfield $n^\mu(x)$ on the bag-boundary $\cmb$.
We will impose {\it local} linear boundary conditions
\eqnl{
\psi(x)=B(x)\psi(x)\mtxt{on}\cmb,}{bag2}
since nonlocal spectral boundary condition, as introduced
and discussed in \cite{AP}, respect the
axial flavour symmetry and probably would
lead to a vanishing condensate in the multiflavour case.
The local boundary conditions must
be compatible with both gauge- and vector-flavour
symmetry which means that $B$ must be a singlett under the
corresponding transformations.
Hence it ought to be in the center of these transformations. \par
The surface integral in \refs{bag1} vanishes if
\eqnl{
B^\dagger\gan B=-\gan \mtxt{and we may assume}B^2=Id.}{bag3}
We shall choose the following one-parametric solution\footnote{Expanding
$B$ in a basis of the Clifford algebra
the general solution in even $d\es 2n$ is found to be
\eqnn{
B_{\theta,\xi}=i\gff\gan\exp\Big(-\theta\gff e^{i\xi\gan}\Big)
\exp\big(-i\xi\gan\big)\times \cc_f\times \cc_c,}
with center elements $\cc$, and depends on two real parameters
$\theta$ and $\xi$.}
\cite{bc2}
\eqnlb{
\psi=B_\theta\psi\mtxt{on}\cmb\quad\mtxt{with}
B_\theta=i\gff e^{\theta\gff}\gan\times I_f\times I_c.}{bag4}
Here $\gff=(-i)^n \gam_0\gam_1\cdots \gam_{d-1}$ is the generalization
of $\gam_5$ which always exists in even dimensions.
We shall choose $\gff=\hbox{diag}(1_n,-1_n)$, i.e. a chiral
representations in which the hermitean $\gam_\mu$ are off-diagonal.
In the following we shall not spell out the trivial action of $B_\theta$
in flavour and colour space as we did in \refs{bag4}.\par
When the 'quarks' are reflected from the bag boundary they
may change their chirality \cite{CDG} which means that
the boundary conditions break the axial-flavour symmetry
\eqnn{
\psi\longrightarrow e^{\gff A}\psi,\mtxt{where}
e^{iA}\in SU(N_f).}
We shall see that $\theta$ in \refs{bag4} plays a similar role as the
$\theta$-parameter in $QCD$.\pan
Let us now derive some properties of the Dirac operator
in an arbitrary external gauge field. The results will
be used later on.
\section{On the Spectrum of the Dirac Operator in a Bag}
The Dirac equation for fermions confined to a bag and subject to the
bag boundary conditions,
\eqnl{
i\di\psi_m(\theta)=\lam_m(\theta)\psi_m(\theta),\qquad
B_\theta\psi_m(\theta)=\psi_m(\theta)\vert_{\cmb},}{fh1}
possesses a discrete spectrum $\{\lam_n\}$.
Unlike the non-zero eigenvalues on a sphere or torus
the eigenvalues do not come in pairs $\lam_n,-\lam_n$. The
reason is that $\psi_n$ and $\gf\psi_n$ can not both obey the
bag boundary conditions. Below
we prove that $i\di$ possesses no zero modes, display how the eigenvalues
and -modes transform under the parity operation and
derive a boundary Hellmann-Feynman formula for the $\theta$-variation
of the eigenvalues.
\subsection{Absence of fermionic zero modes.}
By explicit mode-analysis Balog and Hrasko have shown \cite{bc2} that
in a $2$-dimensional spherical bag $i\di$ possesses no zero modes
which obey the bag boundary conditions \refs{bag4}. Here we shall extend
their result to arbitrarily shaped even-dimensional bags. Indeed, if
there would be a zero mode $\psi$ then we would arrive at the contradiction
\eqnn{
0=(\gff\psi,i\di\psi)-(i\di\gff\psi,\psi)=i\oint \psi^\dagger \gff\gan\psi
=\oint \psi^\dagger e^{-\theta\gff}\psi >0.}
Here we used that as elements of the Clifford algebra $\gff$ and
$\di$ anticommute\footnote{This is not true on the Hilbertspace defined by
\refs{bag4} since $\gff$ does not commute with the boundary conditions. But
this is not needed to arrive at the contradiction.}, the identity \refs{bag1}
and the boundary conditions \refs{bag4} which a possible zero mode
would have to obey.

\subsection{Parity transformations.}
Here we study how the eigenvalues $\lam_m(A,\theta)$ in \refs{fh1}
change under parity transformations of the gauge field
\eqngrl{
A_0(x)&\longrightarrow& \tilde A_0(x)=A_0(\tilde x)\mtxt{,}
\tilde x=(x^0,-x^i)}
{A_i(x)&\longrightarrow& \tilde A_i(x)=-A_i(\tilde x).}{par1}
First we notice that the transformed modes
\eqnl{
\tilde\psi_m(x)=\gff\gamma_0\psi_m(\tilde x)}{par2}
solve the Dirac equation with potential
$\tilde A$ and eigenvalues $-\lam_m$. Second, if $\psi_m$
obeys the boundary condition \refs{bag4} then $\tilde\psi_m$
does, but with $\theta$ replaced by $-\theta$. In other words
\eqnl{
\lam_m (\tilde A,\theta)=-\lam_m (A,-\theta)}{par3}
and this property will constrain the fermionic determinants
and Green's functions.
\subsection{A boundary Hellmann-Feynman formula.}
The Hellmann-Feynman theorem \cite{HelFeyn} relates the
infinitesimal variation of an eigenvalue with the expectation value
of the infinitesimal variation of the operator in the
corresponding normalized eigenstate. Here
we derive a similar formula for the variation of the eigenvalues
$\lam_m$ when the parameter $\theta$ entering the
boundary conditions is varied. \par
To continue we choose the eigenfunctions $\psi_m(\theta)$
in \refs{fh1} to be orthornormal for all values of $\theta$.
The $\theta$-variation of the eigenvalues is then simply
\eqnl{
{d\ov d\theta}\lam_m\equiv\lam_m^\pr=
(\psi_m^\pr,i\di\psi_m)+(\psi_m,i\di\psi_m^\pr)
=i\oint\limits_\cmb\psi_m^\dagger\gan\psi_m^\pr,}{fh2}
where we made use of \refs{bag1}.
The last expression depends only on the eigenmodes restricted
to the bag boundary and there the boundary conditions \refs{bag4}
imply $\psi_m^\pr=\gff\psi_m+B\psi_m^\pr$.
Using the boundary conditions once more, together with the first
formula in \refs{bag3}, we arrive at
\eqnn{
i\psi_m^\dagger\gan\psi_m^\pr=i\psi_m^\dagger\gan \gff\psi_m
+i\psi_m^\dagger\gan B_\theta\psi_m^\pr=i\psi_m^\dagger \gan\gff\psi_m
-i\psi_m^\dagger\gan\psi_m^\pr}
and this can be solved for $i\psi_m^\dagger\gan\psi_m^\pr$.
Inserting the resulting expression into \refs{fh2} finally yields
\eqnlb{
{d\ov d\theta}\lam_m={i\ov 2}\oint \psi_m^\dagger\gan\gff\psi_m
=-\lam_m (\psi_m,\gff\psi_m),}{fh3}
where once again we made use of \refs{bag1}. Eq. \refs{fh3} is the analog of
the Hellmann-Feynman formula and exhibits how $\lam_m$ changes
if the boundary conditions are varied.
\section{The Fermionic Green's Functions}
When calculating correlators of 'quark' fields in a bag
one needs in an intermediate step the Green's function $S^\theta$
of the Dirac operator in an arbitrary background field.
This Green's function must obey
\eqnl{
i\di S^\theta (x,y;A)=\delta(x,y)\mtxt{,}
B_\theta (x)S^\theta (x,y;A)=S^\theta (x,y;A)\vert_{x\in\cmb},}{green1}
and the adjoint relations with respect to $y$. Since
\eqnn{
\di e^{\ha\theta\gff}=e^{-\ha\theta\gff}\di\mtxt{and}
B_\theta e^{\ha\theta\gff}=e^{\ha\theta\gff}B_0}
its $\theta$-dependence is easily found to be
\eqnl{
S^\theta=e^{\ha\theta\gff}\,S^0\, e^{\ha\theta\gff}
=\pmatrix{e^\theta S_{++}^0& S_{+-}^0\cr
          S_{-+}^0& e^{-\theta} S_{--}^0\cr },}{green2}
where the subscripts indicate the chiral projections,
for example $S_{++}=P_+SP_+, P_{\pm}=\ha(1\pm\gff)$. Note that
the $\theta$-dependent diagonal entries $S_{\pm\pm}$ lead
to chirality violating amplitudes and may therefore
trigger a chiral symmetry breaking. Also note that when
we parity-transform the eigenvalues and eigenmodes
of the Dirac operator in the spectral resolution of the Green's function
according to (\ref{par1}-\ref{par3}) we conclude that
\eqnl{
S^\theta(x,y;A)=-\gam_0\gff S^{-\theta}(\tilde x,\tilde y;\tilde
A)\gff\gam_0.}{green3}
This property will relate different correlators
in the fully quantized theories.\par
Next we derive some explicit expressions for $S^\theta$
in spherical bags when the
gauge field is switched off. These free Green's functions are needed
in perturbative expansions for small couplings and/or
small bags. For the explicit calculation it is useful to observe that in a
spherical bag $B_\theta$ commutes with the total angular momentum,
\eqnn{
J_{\mu\nu}={1\ov i}\big(x_\mu\pa_\nu\!-\!x_\nu\pa_\mu\big)+{1\ov 4i}
[\gam_\mu,\gam_\nu]\equiv M_{\mu\nu}+\Sigma_{\mu\nu},}
so that the free Green's functions are rotationally invariant,
\eqnn{
US^\theta (Rx,Ry;0)U^{-1}=S^\theta (x,y;0),\mtxt{where}
U\gam^\mu U^{-1}=R(U)^\mu_{\;\nu}\gam^\nu,}
and only depend on the rotationally invariant quantities
$(\gam,x),(\gam,y),x^2,y^2$, $(x,y)$ and the bag-radius $R$. If we
continue to Minkowski spacetime then $\cm$ becomes the
interior of a hyperboloid and all non-vanishing correlators
would be Lorentz invariant.\par
We may compute the free Green's functions either by angular-momentum
decomposition or by applying the mirror charge method.
We found that their chirality conserving off-diagonal terms
are just those on the infinite
spacetime\footnote{This is a particular property of the
spherical geometry. For instance, on the torus the off-diagonal
terms are modified.} but also that
they contain chirality violating diagonal terms. The final
result in $d\es 2n$ dimensions reads
\eqnlb{
S^\theta (x,y;0)=S_0(x,y)+{\Gamma(n)\ov 2R\pi^n}
\gff\,e^{\theta\gff}
{R^2-(x,\gam)(y,\gam)\ov \big(R^2-2xy+{x^2y^2\ov R^2}\big)^n}
,}{green4}
where
\eqnl{
S_0(x,y)={\Gamma(n)\ov 2i\pi^n}{(x-y,\gam)\ov \vert x\!-\!y\vert^d}}{green4a}
is the free Green's function in $d$-dimendional Euclidean
spacetime. Beeing the Green's functions of a selfadjoint operator
they fulfill the reality condition $S^{\theta\dagger}(x,y)\es
S^\theta (y,x)$. In $2$ dimensions \refs{green4} has been
derived earlier in \cite{bc2}. \par
Note that the $\theta$-dependent chirality violating entries
$S^\theta_{\pm\pm}$ are regular at all interior points and vanish
if the bag size tends to infinity. For example, at the center of the bag
\eqnl{
S^\theta_{\pm\pm}(0,0;0)=\pm{e^{\pm\theta}\ov 2\pi^n}\Gamma(n)R^{1-d}
\longrightarrow 0 \mtxt{for}R\to \infty.}{green5}
They become singular only if $x$ and $y$ both approach the
boundary and each other since then the mirror charge comes close
to $\pa\cm$,
\eqnl{
S^\theta_{\pm\pm}\big(\vert x\vert\es R,y\es (1\!-\!\eps)x;0\big)
\sim \eps^{1-d}.}{green6}
The Green's functions of the squared Dirac operator,
\eqnl{
G^\theta(x,y;A)=\langle x\vert {1\ov -\di^2}\vert y \rangle}{green7}
obey the same boundary conditions as $S^\theta$ and in
addition
\eqnl{i\di G^\theta(x,y;A)=S^\theta (x,y;A).}{green7a}
They transform under the parity operation as
\eqnl{
G^\theta(x,y;A)=\gam_0\gff G^{-\theta}(\tilde x,\tilde y;\tilde A)\gff\gam_0.}
{green8}
After some manipulations we arrived at the following
explicit formulae:
\eqnl{
G^\theta(x,y;0)=G^D(x,y)-C_\theta (x)F(x,y)C_\theta^\dagger (y),}{green9}
where the Dirichlet Green's functions $G^D$ are constructed from
the infinite spacetime Green's functions
\eqnl{
G_0(x,y)=-{1\ov 2\pi} \log \mu\vert x-y\vert\mtxt{resp.}
G_0(x,y)={\Gamma(n-1)\ov 4\pi^n}\,\vert x-y\vert^{2-d}}{greena}
in $2$ and more than $2$ dimensions, respectively, by the
mirror charge method and are found to be
\begin{eqnarray}
G^D(x,y)&=&G_0(x,y)-\big({R^2\ov x^2}\big)^{n-1}G_0(x^\pr,y)
\;\qquad\hbox{for }d>2\label{greenbb}\\
G^D(x,y)&=&-{1\ov 2\pi}\log\Big({R\ov\vert x\vert}{\vert x-y\vert\ov \vert
x^\pr-y\vert}\Big)\quad\qquad\qquad\hbox{for }d=2.\label{greenb}
\end{eqnarray}
Here $x^\pr \es R^2 x/x^2$ denotes the mirror point of $x$.
We have introduced the functions
\eqnn{
C_\theta (x)=1+iR\gff e^{\theta\gff}{(\gam,x)\ov x^2}.}
and
\eqnn{
F(x,y)={i(\gam,x)\ov R}\oint\limits_\cmb \!\!S_0(x,z)G_0(z,y)d\Omega(z)
=\oint\limits_\cmb \!\!G_0(x,z)S_0(z,y)d\Omega(z){(\gam,y)\ov iR},}
where the $z$-integration extends over the bag-boundary.
That $G^\theta$ in \refs{green9} obeys the boundary conditions
is easily verified. To check \refs{green7a} one needs to
use the identity
\eqnn{
\oint S_0(x,z)S_0(z,y)d\om (z)= {\Gamma(n)\ov 2R\pi^n}
{R^2-(\gam,x)(\gam,y)\ov \Big(R^2-2(x,y)+{x^2y^2\ov R^2}\Big)^n}.}
We have calculated $F(x,y)$ in $2$ and $4$ dimensions explicitly.
In $2$ dimensions it reads
\eqnl{
F(x,y)={1\ov 4\pi}\log\Big(1-{\gamma x\gamma y\ov R^2}\Big),}{green13}
and in $4$ dimensions
\eqngrl{
F(x,y)&=&{1\ov 8\pi^2}\Bigg\{
{x^2y^2-(x,y)(\gam,x)(\gam,y)\ov \Delta^{3/2}}
\arctan {\sqrt{\Delta}\ov R^2-(x,y)}}
{\quad&-&{1\ov \Delta}
{\big[R^2-(x,y)\big]x^2y^2-\big[R^2(x,y)-x^2y^2\big](\gam,x)(\gam,y)
\ov R^4-2R^2(x,y)+x^2y^2}\Bigg\},}{green12}
where $\Delta\es x^2y^2-(x,y)^2$.
\section{The Fermionic Determinant in a Bag}
In this section we shall compute the $\theta$-dependence of the
fermionic determinants.
We shall see that the interaction of the fermions with
the bag-boundary induces a $CP$-violating $\theta$ term
in the effective action for the gauge bosons.\par
The Dirac operator and boundary conditions are both flavour
neutral and hence the determinants are the same for all flavours and is
suffices to study the $1$-flavour models.
For the explicit calculations we employ the gauge invariant
$\zeta$-function definition of the determinants \cite{zetaf}
\eqnl{
\log\dett(i\di)\equiv \ha\log\dett (-\di^2)=-\ha {d\ov ds}\zeta_\theta(s)
\vert_{s=0}}{det1}
and calculate their $\theta$-dependence with the help of the
boundary Hellmann-Feynman formula \refs{fh3}.
Denoting the eigenvalues of $-\di^2$ by $\mu_m$, the
$\zeta$-function is defined by
\eqnl{
\zeta_\theta(s)=\sum_m \mu_m^{-s}(\theta)={1\ov \Gamma(s)}
\int\limits_0^\infty\,dt\, t^{s-1}{\tr}_\theta \,e^{t\di^2}
\mtxt{,}\Re (s)>{d\ov 2}}{det2}
and its analytic continuation to $\Re (s)\leq d/2$. Using \refs{fh3}
and the fact that $i\di$ possesses no zero modes, so that a partial
integration with respect to $t$ is justified, the $\theta$-variation
of $\zeta_\theta$ is found to be
\eqnl{{d\ov d\theta}\zeta_\theta(s)=
{2s\ov \Gamma(s)}\int t^{s-1}\tr_\theta\, e^{t\di^2}\gff.}{det3}
Now we can insert the asymptotic small-$t$ expansion of the
heat kernel of $-\di^2$ \cite{SdW} to arrive at the general result
\cite{zetaf1,zetaf2,BVW}
\eqnl{
{d\ov d\theta}\log{\dett}(i\di)=-{1\ov (4\pi)^n}\int\limits_\cm \tr
a_n(\gff )-{1\ov (4\pi)^n}\oint\limits_\cmb \tr b_n(\gff)}{det4}
which holds in an arbitrary $2n$-dimensional bag.
Here the $n$'th ($n\es d/2$) Seeley-deWitt coefficients in the small
$t$-expansion of the heat kernel,
\eqnl{
{\tr}_\theta e^{t\di^2}\phi\sim
{1\ov (4\pi t)^n}\sum_m t^{m/2}\;\tr\Big\{\int a_{m/2}(\phi)+\oint b_{m/2}
(\phi)\Big\}}{det5}
showed up. Unlike the $a_n$ the coefficients
$b_n$ depend on the boundary conditions and thus on $\theta$. \par
For the squared Dirac operator, $\di^2=D^2+\Sigma^{\mu\nu}
F_{\mu\nu}$, that part of the ($\theta$-independent)
$a_n$ which leads to a non-vanishing $\gff$-trace is known in any
dimension \cite{SdW} and inserting it we obtain
\eqnlb{
\log{\dett i\di \ov {\det}_0 i\di}=
{-\theta\ov n!(4\pi)^n}\int\limits_\cm\!\eps_{\mu_1\dots\mu_d}
F_{\mu_1\mu_2}\dots F_{\mu_{d-1}\mu_d}-
\int\limits_0^\theta d\theta^\pr \!\oint\limits_\cmb \tr b_n(\gff),}{det6}
and this formulae are the main results of this section. We see that
the $\theta$ variation is proportional to the
parity-odd instanton number $q$ which is not quantized in a bag.
Our result is in agreement with
\eqnl{
\det i\di(A,\theta)=\det i\di(\tilde A,-\theta)}{det7}
which immediately follows from \refs{par3} and the fact
that the determinant of $i\di$ is defined via
the spectrum of $-\di^2$. This relation means that
parity odd (even) factors in the determinant
are multiplied by functions that are odd (even) in $\theta$
so that the last surface integral in \refs{det6} must be parity odd.
Since the Yang-Mills action is parity even we immediately see that
the measure of functional integration \refs{ein8} satisfies
\eqnlb{d\mu_\theta(A)=d\mu_{-\theta}(\tilde A)}{det8}
which implies that expectation values of parity even (odd)
operators are even (odd) functions of $\theta$.
\par
In particular in $2$-dimensions we have
\eqnl{
\log{\dett i\di\ov {\det}_0  i\di}
=-{\theta\ov 2\pi}\int\tr F_{01},}{ll26}
where we have already anticipated that $\oint\tr b_1(\gff)\es 0$, a fact
that is proven by the heat kernel method in the appendix.
In $4$ dimensions we find
\eqnl{
\log{\dett i\di\ov {\det}_0 i\di}
=-{\theta\ov 2(4\pi)^2}\int\eps_{\mu\nu\al\be}\,
\tr F_{\mu\nu}F_{\al\be}+\oint f_4(\theta,A).}{ll27}
An explicit calculation of surface coefficients like $b_2$ (which
leads to the last surface integral) is not an easy task
\cite{zetaf1}. Contrary to $b_1$ we did not compute it
explicitly. However,
there seems to exist no local polynomial which is parity odd,
gauge invariant and has dimension $-3$ and thus may contribute
to $b_2$. Thus we believe that this surface term is absent as it
is in $2$ dimensions.\par
Since the Dirac operator in a bag is hermitean its determinant is real
and positive. Thus, to make contact with the $\theta$-worlds in $QCD$
\cite{theta} we would have to continue
$\theta$ in \refs{ll27} to $i\theta$. However, when doing this
replacement naively in (\ref{ll26},\ref{ll27}) then
then one runs into
the following apparant paradox: the boundary conditions and
thus the eigenvalues and Green's function of $i\di$ are unchanged if
we replace $\theta$ by $\theta+2\pi in,\;n\in Z$. On the other hand,
the determinant seems not to be periodic since the instanton number is not
quantized. The solution of this apparant paradox is simply
that $\theta$ in (\ref{ll26},\ref{ll27}) should read $\log(e^\theta)$
as is shown in the appendix.
\section{Effective Action in $2$-dimensional Bags}
It has been realized by Polyakov and Wiegmann \cite{PW}
and Alvarez \cite{Al} that the fermionic determinant
on the $2$-dimensional plane may be computed exactly using the
chiral anomaly. Here we shall extend their result to
fermions confined in a $2$-dimensional bag.\par
We shall employ the deformation technique developped in
\cite{sw,zetaf1,BVW} to find the various contributions to the
fermionic determinant. For that we recall that
an arbitrary gauge potential in a two-dimensional bag
(without holes) can always be written as \cite{BVW,Yang}
\eqnl{
A_z\equiv A_0-iA_1=ig^{-1}(\pa_0-i\pa_1)g\equiv ig^{-1}\pa_z g}{def1}
with $g$ from the complexified gauge group $G^c$,
e.g. $g\in GL(n,C)$ for $U(n)$-gauge theories\footnote{On
compact spacetimes without boundaries \refs{def1}
needs some modifications, see \cite{BVW}.}.
Now it is easy to see that
\eqnl{
\di=G^\dagger\fdi G,\mtxt{where}G=
\pmatrix{g^{-1\dagger}&0\cr 0&g\cr }\mtxt{,}
\fdi=\pmatrix{0&\pa_z\cr \pa_{\bar z}&0\cr }}{def2}
and we made the matrix-forms in spinor space explicit.
Note that if we replace $g$ by $gU$, where
$U$ lies in the gauge group $G$, then
\eqnl{G\longrightarrow GU\mtxt{and}
\di\longrightarrow U^{-1}\di U}{def3}
and hence the corresponding gauge potential is just the
gauge-transformed one. The field strength is
\eqnl{F_{01}=-\ha g^\dagger \bar\pa\big(J^{-1}\pa J\big)g^{-1\dagger}
            =-\ha g^{-1}\pa\big(\bar\pa J J^{-1}\big)g,}{def4}
where the gauge invariant field
\eqnl{J=gg^\dagger}{def6}
with values in the coset space $G^c/G$ appeared.
$J$ will play an important role since all gauge invariant
Green's functions depend on the gauge field only via this gauge
invariant field. The Yang-Mills
action reads
\eqnl{
S_{YM}={1\ov 8g^2}\int\tr \bar\pa(J^{-1}\pa J)\bar\pa(J^{-1}\pa J).}{def5}
\par
Let us now introduce a $\tau$-dependent family $g(x,\tau)$
which interpolates between the identity and the field $g(x)$ as
\eqnl{
g(x,0)=I\mtxt{,}g(x,1)=g\mtxt{and}
{d\ov d\tau}g(\tau)\equiv \dot g(\tau)=-g(\tau)a(\tau).}{def7}
With \refs{def2} it follows at once that
\eqnl{
\dot\lam_m=\lam_m(\psi_m,(A+A^\dagger)\psi_m)+i\oint
\pd_m\gan A\psi_m,\quad A=\pmatrix{a^\dagger&0\cr0&-a\cr }.}{def8}
To get rid of the annoying surface term we observe that
the gauge potential in \refs{def1} is unaffected by the replacement
\eqnl{
g\longrightarrow \al^{-1}(\bar z)g}{def9}
and we can use this freedom to get rid of this term.
Indeed, we can always find a unique $\al$ such that
$\al(\bar z)\al(z)^\dagger\es gg^\dagger$ on the bag-boundary.
The equivalent $g$ obeys then
\eqnl{
gg^\dagger\vert_\cmb\equiv J\vert_\cmb=I\Longleftrightarrow
G^{-1}B_\theta G=B_\theta\mtxt{on}\cmb.}{def10}
Imposing the first condition for all $\tau$ implies that on the
bag boundary $a+a^\dagger\es 0$ or that $A$ is the identity
in spinor space. Then the surface term in \refs{def8} vanishes
on account of the bag boundary conditions.
The second condition is just the statement that the
$G$-transformation \refs{def2} is compatible with the bag
boundary conditions so that the Green's function is related to
the free one\footnote{
we use the same symbol $S^\theta (x,y;0)$ independently on
whether the free Green's function \refs{green4} is tensored with
the identities in flavour- and/or colour space or not. The local
meanings should be clear from the context.}, \refs{green4}, as
\eqnl{
S^\theta (x,y;A)=G^{-1}(x)S^\theta (x,y;0)G^{-1\dagger}(y).}{def11}
In the following we assume \refs{def10} to hold
for all $\tau$ so that the whole deformation \refs{def7}
is compatible with the boundary conditions.

Now we can apply the wellknown deformation techniques for
the $\zeta$-function defined determinant \cite{zetaf1,BVW} and obtain
\eqnl{
{d\ov d\tau}\log\det i\di={1\ov 4\pi}\int\limits_\cm
\tr a_1(A+A^\dagger)+{1\ov 4\pi}\oint\limits_{\cmb}\tr
b_1(A+A^\dagger).}{def12}
Here $A$ and the Seeley-deWitt coefficients $a_1,b_1$ of the
$\tau$-deformed Dirac operator are to be inserted.
The volume coefficient $a_1$ is wellknown \cite{SdW},
\eqnl{
\int a_1(\phi)=\int F_{01}\gff\phi}{def13}
contrary to the surface coefficient $b_1$. We have calculated $b_1$
via the heat-kernel in the appendix and up to purely geometric
terms, which cancel in expectation values, the result is
\eqnl{
\oint b_1(\phi)=\ha\oint \Big\{1-{\log e^\theta\ov\sinh(\theta)}
\pmatrix{e^\theta&-1\cr -1 &e^{-\theta}\cr}\Big\}\pa_n \phi.}{def14}
Note that for a constant function $\phi$ the surface Seeley-deWitt
coefficient $b_1(\phi)$ vanishes, and we have used this
fact earlier in deriving \refs{ll26}. Note, however, that
although $A+A^\dagger\es 0$ on $\cmb$ the last surface
integral in \refs{def12} does not vanish, since $\tr b_1(\phi)$
contains the normal derivatives of $\phi$ at the boundary.\par
Inserting \refs{def14} into \refs{def12} we end up with the exact formula
\eqnl{
\log{\dett i\di\ov \dett i\fdi}={1\ov 2\pi}\int\limits_0^1
d\tau\Big\{\int\limits_\cm\tr F_{01}(a+a^\dagger)
-{\theta\ov 2}\oint\limits_\cmb \tr\pa_n(a+a^\dagger)\Big\}.}{def15}
To continue we express $a$ and $F_{01}$ in terms of $g$ and
its derivatives and find
\eqnn{
\log{\dett i\di\ov \dett i\fdi}=-{1\ov 4\pi}\int\limits_0^1 d\tau\Big\{
\int\limits_\cm\tr\Big(J^{-1}\pa J\bar\pa(J^{-1}\dot J)\Big)
-\theta\oint\limits_\cmb \tr\pa_n\big(J^{-1}\dot J\big)\Big\}.}
The $\tau$-integral of the volume term can be calculated
in the same way as on the infinite plane\footnote{the various
partial integrations needed to arrive at the result are allowed
if one takes into account that $J$ is the identity on the bag-boundary}
and leads to the Wess-Zumino action \cite{BVW,KM}.
That of the surface term is easily found since $\pa_\tau\tr\bar
\pa(J^{-1}\pa J)=\lap(J^{-1}\dot J)$. Hence we arrive at the following
explicit answer for the fermionic determinant in a bag:
\eqngrl{
\log{\dett i\di\ov \dett i\fdi}=&-&{1\ov 8\pi}
\int\limits_\cm\tr\Big(J^{-1}\pa J J^{-1}\bar\pa J\Big)
+{i\ov 12\pi}\int\limits_{\cal Z}\tr \big(J^{-1}d_3J\big)^3}
{&+&{\theta\ov 4\pi}\int\limits_\cm\tr\bar\pa\big(J^{-1}\pa J\big).}{def16}
In the Wess-Zumino term in the middle on the right hand
side $J\es J(x,\tau)$ and thus ${\cal Z}\es \cm\times [0,1]$
is the finite cylinder over the bag.
We recall that the deformation is subject to the
boundary-, initial- and final conditions
\eqnl{J(x\in\cmb,\tau)=I\mtxt{,}J(x,0)=I\mtxt{and} J(x,1)=J(x).}{def17}
As for the last surface term in \refs{def16} we see
immediately that for semisimple gauge groups it vanishes, since
$J^{-1}\pa J$ lies in the complexified gauge algebra.
Also note that this term is equal to $-\theta/2\pi\int\tr F_{01}$
so that our result is indeed compatible with \refs{ll26}.
Also, for $J\es J_1J_2$ it becomes the sum of such terms for the individual
fields $J_i$. This means that the wellknown Polyakov-Wiegman
identity \cite{PW}, which relates the determinant belonging
to $J\es J_1J_2$ with those of $J_1$ and $J_2$,
\eqngrl{
\log{\dett i\di(J_1J_2)\ov \dett i\fdi} &=&\log{\dett i\di (J_1)
\ov \dett i\fdi}+\log{\dett i\di (J_2)\ov \dett i\fdi}}
{&-&{1\ov 4\pi}\int\limits_\cm
\tr\Big(J_1^{-1}\pa J_1\bar\pa J_2J_2^{-1}\Big).}{def16a}
still holds in a bag.\par
Let us now suppose that $G=U(1)\times SU(N_c)$. The results for
this particular case will be important when we calculate
mesonic current correlators and chiral condensates.
We represent the gauge potential $A\es \tilde A+\hat A$ as in
\refs{def1} and factorize the $U(1)$ field, that is we set
$g\es \tilde g\hat g$. We parametrize the $U(1)$-part as
$\tilde g=e^{-e\varphi-ie\lam}$, where $e$ is the electric charge,
and then
\eqnl{
A_\mu=\tilde A_\mu+\hat A_\mu=-e\eps_{\mu\nu}\pa_\nu\varphi+e\pa_\mu\lam
+\hat A_\mu\mtxt{and} F_{01}=e\lap\varphi+\hat F_{01}.}{def18}
Repeating the above analysis for the deformation
\eqnn{
J(x,\tau)=e^{-2e\varphi(x) \tau}\hat J(x,\tau)\mtxt{with}
\varphi\vert_\cmb=0\mtxt{and}\hat J(\tau)\vert_\cmb=I,}
or equivalently applying the Polyakov-Wiegman identity
to $J\es e^{-2e\varphi}\hat J$, shows that the
determinant \refs{def16} factorizes,
\eqnlb{
\dett i\di=
e^{-{N_c\ov 2\pi}[e^2\int\pa\varphi\bar\pa\varphi+\theta e\oint\pa_n\varphi]}
\det i\hat\di,}{def19}
where the last determinant is $\theta$-independent. The same
happens then for the functional measure for the Euclidean
gauge fields
\eqnl{
d\mu_\theta (A)=d\mu_\theta (\tilde A)\;d\mu (\hat A)=
{e^{-\Gamma_\theta [\varphi]}\ov \tilde Z _\theta}\cd \tilde A\;\;
{e^{-\Gamma[\hat A]}\ov \hat Z}\cd \hat A.}{def20}
Here we introduced the $\theta$-dependent effective action
for the $U(1)$-gauge potential $\tilde A$ and the $\theta$-independent
one for the $\hat G$-gauge potential $\hat A$. For the
$N_f$-flavour model with flavour-independent
$U(1)$-charge $e$ and $\hat G$-coupling constant $g$ they read
\eqngrl{
\Gamma_\theta[\varphi]&\es&
{N_c\ov 2}\Bigg\{\int (\lap\varphi)^2-m^2_\eta\int \varphi\lap\varphi+
{e\theta N_f\ov \pi}\oint \pa_n\varphi\Bigg\}}
{\Gamma[\hat A]&\es&S_{YM}[\hat A]+{N_f\ov 8\pi}\!\int\limits_\cm
\tr\big(\hat J^{-1}\pa \hat J \hat J^{-1}\bar\pa \hat J\big)
-{iN_f\ov 12 \pi}\!\int\limits_{\cal Z}\tr\big(\hat J^{-1}d_3\hat J\big)^3.}
{def21}
Note that due to the wellknown Schwinger mechanism the mass
\eqnl{
m^2_\eta =N_f{e^2\ov \pi},}{def21a}
which is the analog of the $\eta^\pr$-mass in $QCD$, has
been induced in the abelian subsector of the theory.
\section{Correlation Functions of Mesonic Currents}
Fermionic correlation functions are gotten from the generating functional
\refs{ein1}, which in a bag simplifies to
\eqnl{
Z[\eta,\bar\eta]=\int d\mu_\theta (A)\;
e^{\int \eta^\dagger(x)S^\theta (x,y;A)\eta(y)},}{corr1}
by functional differentiation with respect to the grassmann
valued sources. Here $d\mu_\theta$ is the measure of functional integration
\refs{ein8} and we recall that the fermionic Green's function $S^\theta$ is
the identity in flavour space. Let $\cc=\cs\otimes \cf\otimes I_c$ be
a numerical matrix which acts trivial in colour space. Then we
obtain for the gauge invariant connected two- and four-point functions
in a fixed background field
\eqngrrl{
&&\<\pd (x)\cc\psi (x)\>_A=-\tr \cf\;\tr\cs\, S^\theta(x,x;A)}
{&&\<\pd(x) \cc_1\psi(x)\pd (y)\cc_2\psi(y)\>_{A,c}}
{&&\qquad\qquad=-\tr \cf_1 \cf_2\;\tr\big[\cs_1 S^\theta
(x,y;A)\cs_2 S^\theta (y,x;A)\big],}{corr4}
where it is understood that the first traces are in flavour space
and the second ones in spinor- and colour space.
\paragraph{Vector currents}
The $2$-point functions of the mesonic vector- and pseudovector currents
\eqnl{
j^\mu_\cf=\pd\gam^\mu\cf\,\psi\mtxt{and}
j^{5\mu}_\cf=\pd\,\bar\gam\gam^\mu\cf\,\psi
=i\eps_{\mu\nu}j^\nu_\cf}{vc1}
will already shed some light on the particle spectrum of $2$-dimensional
gauge theories. We obtain the following formal expressions for the
connected $1$ and $2$-point functions
\eqngrl{
\<j^\mu_\cf(x)\>_A &=&-\tr\cf\;\tr \gam^\mu S^\theta (x,x;A)}
{\<j^\mu_{\cf_1}(x)j^\nu_{\cf_2}(y)\>_{A,c}&=&
-\tr\cf_1\cf_2\;\tr \gam^\mu S^\theta(x,y;A)\gam^\nu S^\theta(y,x;A).}{vc3}
In $2$ spacetime dimensions the Green function $S^\theta$
is given by \refs{def11} and \refs{green4}.
When inserting the explicit form (\ref{def11},\ref{green4}) of $S^\theta$
one notices that the gauge field and $\theta$-parameter both
drop in these expectation values. In principle one would have
to regularize the currents, e.g. by a gauge invariant point
splitting prescription and this may reintroduce
a gauge field and $\theta$-dependence. However, by noticing
that the mesonic currents couple to the abelian gauge potential
$\tilde A_\mu$ in \refs{def18} we can calculate the
regularized connected correlators in a fixed background as
\eqngrl{
&&\<j^\mu_\cf (x)\>_A=
{\tr\cf\ov e}{\delta\log\det i\di \ov\delta \tilde A_\mu(x)}
={N_c\tr\cf\ov \pi}\eps_{\mu\nu}\Big(e
\pa^\nu\varphi+{\theta\ov 2}\delta(r\!-\!R)n^\nu\Big),}
{&&\<j^\mu_{\cf}(x)j^\nu_{\cf}(y)\>_{A,c}
={\tr\cf^2\ov e^2}{\delta^2 \log\det i\di
\ov \delta \tilde A_\mu(x)\delta \tilde A_\nu(y)}
=-{N_c\tr\cf^2\ov\pi}{\cal P}^{\mu\nu}(x,y).}{vc5}
All higher connected correlators vanish. In deriving \refs{vc5}
we have factorized the flavour dependence by diagonalizing
$\cf$ so that the determinants are those of the one-flavour model.
The last equalities follow from the explicit dependence of
$\det i\di$ in \refs{def19} on the field $\varphi$
and the decomposition of $\tilde A_\mu$ in \refs{def18}.
${\cal P}^{\mu\nu}$ projects onto the transversal degrees of
freedom and is consistent with the boundary conditions,
\eqnl{
{\cal P}^{\mu\nu}(x,y)=\pi\tr\gam^\mu S^\theta(x,y;0)\gam^\nu
S^\theta (y,x;0)=\eps^{\mu\al}\eps^{\nu\beta}\pa_{x^\al}
\pa_{y^\beta}G^D(x,y).}{vc5b}
Here $G^D(x,y)$ is the Dirichlet Green's function of $-\lap$,
see \refs{greenb}.
Since $\varphi\es 0$ on $\cmb$ the current normal to $\cmb$
vanishes and no $U(1)$-charge is leaking through
the boundary as required by the boundary
conditions on the 'quark' fields. Furthermore, our result is compatible
with vector flavour symmetry and the axial vector anomaly,
\eqnl{
\pa_\mu\<j^\mu_\cf\>_A=0\mtxt{and}
\pa_\mu\<j^{5\mu}_\cf\>_A={\tr\cf\ov i\pi}\,\tr_c\Big\{
e F_{01}+{\theta\ov 2}\delta^\pr (r-R)\Big\}.}{vc4a}
Note that the nonabelian part $\hat A$ of the gauge potential
has completely disappeared in the above formulae.
Since we know all correlators in an arbitrary gauge field and since
those only depend on the abelian part of the gauge potential
the averaging over the gauge fields reduces to that
in the multi-flavour Schwinger model.
Here we may use the results in \cite{GS}, up to some modification
due to the presence of the bag boundary.
Let us choose a trace-orthonormal basis $T_a,\,
a=2,3,\dots,N_f^2$ of $SU(N_f)$, together with the identity
in flavour space which we denote by $T_1$.
The correlators of the associated currents
$j^\mu_a=\bar\psi\gam^\mu T_a\psi$ are reproduced by the
generating functional
\eqngrl{
&&\<\exp\Big(\int j^\mu_a b^a_\mu\Big)\>
=\exp\Bigg\{-{N_c\ov 2}\Big[m^2_\eta\int b^1_\mu(x)
{\cal P}^{\mu\nu}_{m_\eta}(x,y)b^1_\nu(y)}
{&&\quad +{m^2_\eta\ov N_f}\sum\limits_2^{N_f^2}b^a_\mu (x)
{\cal P}^{\mu\nu}(x,y)b^a_\nu(y)+{e\theta N_f\ov \pi}\int I_e(r,R)
\eps^{\mu\nu}\pa_\mu b^1_\nu\Big]\Bigg\},}{vc4b}
where we introduced the function
\eqnl{
I_e(r,R)={I_0(m_\eta r)\ov I_0(m_\eta R)}.}{lll7b}
The projector ${\cal P}^{\mu\nu}_m$ onto the transverse massive
vector-particles is derived from the massive Green function
$G_m^D$ in \refs{lll8} in the same way as ${\cal P}^{\mu\nu}$ was derived
from $G^D$ in \refs{vc5b}. Actually, the generating functional for the
currents in the Cartan subalgebra can be calculated directly
since the associated fermionic determinant is calculable. The identities
needed to prove that the generating functional \refs{vc4b} yields the correct
current correlators are derived in the next section, see for example
\refs{lll7a}.\par
Now it is easy to bosonize the mesonic currents, since the bosonization
is identical to that of the multi-flavour Schwinger model \cite{GS},
up to boundary terms. One finds that the generating functional
for all currents can be rewritten as
\eqnl{
\<\exp\Big(\int j^\mu_a b_\mu^a\Big)\>=
\<\exp\Big(i\int \eps^{\mu\nu}\pa_\nu\,\varphi_a b^a_\mu\Big)\>_B,}{vc6}
where the Gaussian measure for the $N_f^2$-bonsonic fields $\varphi_a$
has the action
\eqnn{
B[\varphi]={1\ov 2N_c m_\eta^2}\Big[\int \varphi_1
\big(\!-\!\lap+m^2_\eta\big)\varphi_1
-N_f\sum_2^{N_f^2}\int\varphi_a \lap\varphi_a\Big]
+{i\theta\ov e}\oint \pa_n\varphi_1.}
We recovered the wellknown bosonization rule $j^\mu_a\to i\eps^{\mu\nu}
\pa_\nu\varphi_a$, where the field $\varphi_1$ belonging to
the $U(1)$-current $\bps \gam^\mu\psi$ has mass
$m_\eta$ and the remaining $N_f^2-1$ pseudo-scalar fields are massless.
What we have shown is that $2$-dimensional multi-flavour $U(N_c)$
gauge theories contain one massive and $N_f^2-1$ massless pseudoscalar
'mesons'. For $G\es SU(N_c)$ the massive 'meson' is absent.
\section{Chiral Symmetry Breaking in $2d$-Gauge Theories}
We begin with calculating the chiral condensate of the
$N_f$-flavour Schwinger model \cite{Schwing,GS}
enclosed in a spherical bag. As an application we
derive an upper bound for the condensate in $SU(N_c)$
gauge theories and prove that for $N_c<\infty$ it
vanishes in the thermodynamic limit. On the other hand,
for the abelian projected non-abelian theories
we calculate the $R$-dependence of the condensate explicitly
and show that in the limit $N_c\to \infty$ a
'quark' condensate is generated which remains when
$R\to \infty$.\par
The $u\es\psi_1$-'quark' condensate is the particular $2$-point-function
\refs{corr4} with $\cs=P_+$ and $\cf_{ab}\es\delta_{a1}\delta_{b1}$. Inserting
the explicit form of the Green function $S^\theta$ we arrive at
\eqnl{
\< \bar uP_+u\>(x)=-{e^{\theta}\ov 2\pi R}{1\ov 1-r^2/R^2}
\int d\mu_\theta(A)\;\tr J(x)}{corr3}
and it remains to calculate the average of the colour trace of the
gauge invariant field $J$.
For $2$-dimensional $SU(N_c)$-gauge theories the measure
$d\mu$ does not depend on $\theta$ and the condensate
is proportional to $e^\theta$. On the other hand, we shall see
that for $U(N_c)$-theories the 'quark' condensates become
$\theta$-independent, up to exponentially small (in $R$) finite
size corrections.
\subsection{Multi-flavour $QED_2$}
When one quantizes multi-flavour $QED_2$ with massless fermions
on $S^2$ \cite{jay} or the torus \cite{sw,Torus} or some other
Riemann surface one finds $\<\bar uP_+ u\>\es 0$.
The same result is found in the geometric Schwinger model \cite{Jo}
which is equivalent to $QED_2$ with $2$-flavours.
The condensate vanishes for the same reason as it does in $QCD$
if one only allows for gauge fields with integer instanton number.
Only for nonzero 'quark'-masses or if one allows for flavour
dependent twisted boundary conditions does one find a nonzero condensate
in finite volumes.
Here we shall show that the $U_A(N_f)$-breaking bag-boundary
conditions also trigger a chiral condensate. No fermionic zero
modes are needed to generate it and actually there are none of them.
The condensate decreases with increasing bag-radius unless
$N_f\es 1$ or the number of colours is infinite.\par
As earlier we choose the parametrization $g\es
e^{-e\varphi-ie\lam}$ (we skip the tilde in this subsection) for
the abelian field so that the functional integral representation
for the $u$-'quark' condensate reads
\eqnl{
\< \bar u P_+ u\>=
- {e^{\theta}\ov 2\pi R}\,{1\ov 1-r^2/ R^2}
{\int \cd A_\mu\, e^{- 2e\varphi(x)-\Gamma_\theta[\varphi]}\ov
\int \cd A_\mu\, e^{-\Gamma_\theta[\varphi]}},}{lll4}
where $\Gamma_\theta[\varphi]$ is the effective action
\refs{def21} for one colour. The Jacobian of the transformation
\refs{def18} from the potential $A_\mu$ (there it was denoted
by $\tilde A_\mu$) to the new fields
$\lam,\varphi$ is field independent and we can
replace $\cd A_\mu$ by $\cd\varphi$ in expectation values
of gauge invariant operators. Also recall that we integrate over those
fields $\varphi$ which vanish on the bag-boundary.\par
The integral \refs{lll4} is Gaussian with source
\eqnn{
j(y)=-2e\delta(x-y)+{e\theta N_f\ov 2\pi}{1\ov r_y}\pa_{r_y}\Big(
r_y\delta(r_y-R)\Big)}
and thus is found to be
\eqnl{
\< \bar u P_+ u\> =
{-e^{\theta}\ov 2\pi R}{1\ov 1-r^2/R^2}\exp\Big\{
{2\pi\ov N_f} K(x,x)+\theta\!\int \!d^2y \lap_y K(x,y)\Big\}.}{lll5}
Here we introduced
\eqnl{
K(x,y)=\< x\vert {1\ov -\lap}\vert y\>-\< x\vert {1\ov
-\lap+m^2_\eta}\vert y\>\equiv G^D(x,y)-G_{m_\eta}^D(x,y),}{lll6}
i.e. the difference between the massless and massive Green's functions
with respect to Dirichlet boundary conditions. In a spherical
bag with radius $R$ $G^D$ has been given in \refs{greenb}
and
\eqngrl{
G_{m}^D(x,y)&=&{1\ov 2\pi}\Big\{K_0(m\vert x-y\vert)}
{&\;\;-&\!\sum\limits_0^\infty \eps_n{K_n(m R)\ov I_n(m
R)}I_n(m r_x)I_n(m r_y)\cos n(\varphi_x-\varphi_y)\Big\},}{lll8}
where $\eps_0\es 1$, $\eps_{n>0}=2$ and $I_n,K_n$ are
the modified Bessel functions. Using the explicit form of
the Green's functions one calculates
\eqnl{
\int\limits_\cm d^2y\,\lap_y G^D(x,y)=-1\mtxt{and}
\int\limits_\cm d^2y\,\lap_y G_m^D(x,y)=-{I_0(mr)\ov I_0(mR)},}{lll7a}
so that
\eqnn{
\<\bar uP_+u\>=-{1\ov 2\pi R}\,{1\ov 1-r^2/R^2}
\exp\Big\{\theta I_e(r,R)+{2\pi\ov N_f}K(x,x)\Big\}.}
The function $I_e$ in the exponent has been defined in \refs{lll7b}.
Inserting the expansion of $K_0$ for small arguments we obtain
\eqnn{
2\pi K(x,x)=\gam+ \log\Big({m_\eta R\ov 2}[1-{r^2\ov R^2}]\Big)
+F_e(r,R),}
where $\gam\es 0.577\dots $ is Euler's constant and we have introduced
the function
\eqnl{
F_e(r,R)=\sum \eps_n{K_n(m_\eta R)\ov I_n(m_\eta R)}I^2_n(m_\eta r),}{lll7c}
Inserting all that we get the following exact formula for the
chiral condensate in multi-flavour $QED_2$ confined in a
bag with radius $R$:
\eqnlb{
\<\bar u P_+ u\>(x)=
-{m_\eta e^\gam\ov 4\pi}\Big({m_\eta R e^\gam\ov 2}
\big[1-{r^2\ov R^2}\big]\Big)^{-1+1/N_f}\;e^{\theta I_e+F_e/N_f}.}{lll9}
The function $F_e$ has the asymptotic expansions
\eqnl{
F_e(r,R)\sim\left\{
\begin{array}{ll}e^{-m_\eta R}&\hbox{for }1\ll m_\eta R\gg m_\eta r\\
-\log \ha m_\eta Re^\gam [1-{r^2\ov R^2}]&
\hbox{for } m_\eta R\ll 1.\end{array} \right. }{lll10}
Thus for large and small bags or equivalently for strong and
weak coupling constant $e$  the condensate simplifies to
\eqnl{
\<\bar u P_+ u\>\sim\left\{
\begin{array}{ll}
-{m_\eta e^\gam\ov 4\pi}\Big(\ha m_\eta Re^\gam
\Big)^{-1+1/N_f}&\hbox{for }1\ll m_\eta R\gg m_\eta r\\
-{e^\theta\ov 2\pi R}\big(1-r^2/R^2\big)^{-1}&
\hbox{for }m_\eta R\ll 1. \end{array} \right. }{lll11}
As expected, for weak couplings and/or small bags the condensate tends to
the chirality violating entry $-S^\theta_{++}(x,x;0)$ of the free Green's
function \refs{green4}.\par
For {\it one flavour} and large bags we recover
the wellknown value for the condensate in the Schwinger model
\cite{Schwinger}
\eqnl{
\<\bar u P_+ u\>=-{m_\eta\ov 4\pi}e^\gam.}{lll12}
We stress that this result has been obtained without doing any
instanton physics. The calculations in a bag are actually much simpler
as compared with those on a torus \cite{sw,Jo,Torus} or sphere \cite{jay},
where a careful treatment of the different instanton sectors
is required to find the result \refs{lll12}.\par
For {\it several flavours} the condensate inside the bag, e.g.
at the center of a large bag,
\eqnl{
\<\bar u P_+ u\>(0)= -{1\ov 2\pi R}
\Big({m_\eta R e^\gam\ov 2}\Big)^{1/N_f}}{lll13}
decreases with increasing bag radius and vanishes in
the thermodynamic limit.\par
The cluster property holds since the $4$-point function
\eqngrrl{
&&\<\bar u(x)P_+u(x)\bar u(y)P_-u(y)\>}
{&&\quad=\<S_{++}^\theta (x,x;A)S_{--}^\theta (y,y;A)-
S_{-+}^\theta(x,y;A)S_{+-}^\theta(y,x;A)\>}
{&&\quad\longrightarrow -\big({m_\eta\ov 4\pi}\big)^2
\Big({m_\eta\ov 2}\vert x-y\vert\Big)^{-2+2/N_f}\;
e^{2\gam/N_f}\mtxt{for }R\to\infty }{lll13a}
tends to the product of the left- and righthanded condensates
for large separations $\vert x-y\vert$.\par
Let us finally prove that in the thermodynamic limit all
fermionic correlators in multi-flavour $QED_2$ become
$\theta$-independent. This follows from the explicit
form of the fermionic Green's function (for $\lam\es 0$)
\eqnn{
S^\theta(x,y;A)=e^{-\gff [e\varphi(x)-\ha\theta]}S^0(x,y;0)
e^{-\gff [e\varphi(y)-\ha\theta]},}
which implies that all correlators are proportional to
\eqnn{
e^{-\theta\sum \al_i}\;\<e^{2e\sum \al_i\varphi (x_i)}\>,}
and from the formula
\eqnn{
\<e^{2e\sum \al_i\varphi (x_i)}\>=
e^{\theta\sum \al_i [1-I_\eps (r_i,R)]}\;e^{2\pi/N_f\sum \al_i K(x_i,x_j)
\al_j}.}
Thus, up to exponentially small finite size corrections $\sim
\exp(\theta I_\eps)$ the $\theta$-dependence cancels in all
fermionic correlators.\par
Let us compare our result with that of Smilga \cite{Smilga1}
who calculated the condensate in multiflavour $QED_2$
for small 'quark' masses. Using bosonization techniques
he found that the mass $\mu$ of the lightest particle
and the 'quark' condensate depend on the
electric charge $e$ and small current quark masses $m$ as
\eqnn{
\mu\sim \big(m_\eta m^{N_f}\big)^{1\ov N_f+1}\mtxt{and}
\<\bps\psi\>\sim \big(m_\eta^2\mu^{N_f-1}\big)^{1\ov N_f+1}}
so that
\eqnl{
\<\bps\psi\>\sim \mu\Big({m_\eta\ov \mu}\Big)^{1/N_f}.}{lll14}
Comparing with \refs{lll13} we see that the bag- and
small quark mass calculations yield the same result if
we identify the mass of the lightest particle in the spectrum
with the inverse radius of the bag. In other words,
small quark masses and bag boundary conditions both
trigger the same condensate if $\mu$ is identified
with $1/R$.\par
In passing we note that the left- and right-handed
condensates are related as
\eqnl{
\<\bar u P_- u\>_\theta=-\<\bar u P_+ u\>_{-\theta}.}{lll15}
This follows from the transformations \refs{det8}
and \refs{green3} under the parity operation. Since the
function $I_e$ in \refs{lll9} vanishes exponentially
with increasing bag radius $R$ (assuming that $r\!\ll\! R$)
we conclude that
\eqnl{
\<\bar u u\>=\<\bar u P_+u\>+\<\bar u P_- u\>=
O\Big(\sinh (\theta e^{-m_\eta R})\Big)}{lll16}
for large bags or in the strong coupling limit.\par
To summarize, up to a phase the thermodynamic limits of the left- and
right-handed condensates in a bag are identical to the
instanton induced condensates in the $1$-flavour model on the torus
or sphere and to the condensates in the multi-flavour models obtained
via perturbative expansion in the small quark masses.
The same is true for the condensate $\<\bar u u\>$ only
for particular values of the parameter $\theta$ in the
$\theta$-world.
\subsection{Multi-flavour nonabelian gauge theories.}
Due to the factorization of the measure for the
gauge bosons, \refs{def20}, the chiral condensate \refs{corr3}
in $U(N_c)$ gauge theories factorizes as
\eqngrl{
\<\bar u P_+u\>_{U(N_c)}&=&
-{e^\theta\ov 2\pi R}{1\ov 1-r^2/R^2}\int d\mu_\theta (\tilde A)e^{-2e\varphi}
\int d\mu (\hat A)\,\tr\hat J}
{&=&-{2\pi R\ov e^{\theta}}\big(1-{r^2\ov R^2}\big)
\;\<\bar u P_+u\>_{U(1)}\;\<\bar uP_+ u\>_{SU(N_c)},}{nagt1}
and thus is proportional to the Schwinger model result times
the $SU(N_c)$ condensate. When calculating the
$U(1)$-condensate one should remember that $\Gamma_\theta[\varphi]$ in
(\ref{def20},\ref{def21}) is $N_c$ times that of the
multi-flavour Schwinger model, so that \refs{lll9} is modified to
\eqnl{
\<\bar u P_+ u\>_{U(1)}=-{e^{\theta I_e}\ov 2\pi R}{1\ov 1-r^2/R^2}
\Big({m_\eta R\, e^{\gam+F_e}\ov 2}
\big[1-{r^2\ov R^2}\big]\Big)^{1/N_c N_f},}{nagt2}
where the functions $I_e$ and $F_e$ have been defined in \refs{lll7b}
and \refs{lll7c}, respectively.
Inserting all that into \refs{nagt1} we find the following
exact relation between the $U(N_c)$ and $SU(N_c)$ condensates:
\eqnl{
\<\bar uP_+u\>_{U(N_c)}=
e^{\theta(I_e-1)}\;\Big({m_\eta R\, e^{\gam+F_e}\ov 2}\big[1-{r^2\ov R^2}\big]
\Big)^{1/N_cN_f}\<\bar uP_+u\>_{SU(N_c)}.}{nagt3}
Using the asymptotic expansion of $F_e$ for small arguments, \refs{lll10},
we see that for $e\to 0$ the $U(N_c)$ result reduces to the $SU(N_c)$ one,
as expected.\par
For the condensates at the center of large bags \refs{nagt3} simplifies to
\eqnl{
\<\bar uP_+u\>_{U(N_c)}=
e^{-\theta}\big[{m_\eta R e^\gam\ov 2}\big]^{1/N_cN_f}
\<\bar uP_+u\>_{SU(N_c)}.}{nagt4}
Assuming that the $U(N_c)$ condensate has a smooth thermodynamic
limit we conclude at once that for
a finite number of colours the quark condensate in $SU(N_c)$ gauge
theories tends to zero as the bag increases at least as
\eqnlb{
\<\bar uP_+ u\>_{SU(N_c)}\leq \hbox{const}\cdot R^{-1/N_cN_f}.}{nagt4a}
Only when we take the limit in which the
number of colours tends to infinity {\it before} we perform the
thermodynamic limit $R\to\infty$ can a quark condensate survive.\par
It would be interesting to see how \refs{nagt4a} is modified for
two-dimensional $QCD$ with adjoint Majorana fermions. Arguments
based on the bosonized representation of the theory imply that
a nonvanishing condensate is generated, even for $N_c\geq 3$
in which case the instantons fail to generate a condensate
\cite{ShifSmil,Smilga3}.
\subsection{Baby-$QCD_2$}
For doing explicit calculations
it is useful to parametrize the $g$-field in \refs{def1}.
We take a parametrization for which the
fermionic determinant becomes local and simple. The price we pay
for the locality is that the Yang-Mills action is not quadratic as it
would be in a gauge like $A_r\es 0$. For simplicity we assume
that $G=SU(2)$, that is we consider the baby-version of
$QCD_2$ \cite{Cole}. For baby-$QCD$ the field $g$ lies in $SL(2,C)$
and in a bag without holes any such $g$ can globally be decomposed as
\cite{Shap}
\eqnl{
g=hU,\mtxt{where}h=\pmatrix{e^{\ha\varphi}& ve^{\ha\varphi}\cr
0&e^{-\ha\varphi}\cr}\mtxt{and} U\in SU(2).}{baby1}
Here $U$ contains the pure gauge part of the potential
and cancels in expectation values of gauge invariant
operators\footnote{The gauge field measure is discussed below}.
The condition \refs{def10} means that $\varphi$ and $v$
both vanish on the bag boundary.\par
Now we can apply the Polyakov-Wiegman identity \refs{def16a}
with $J\es hh^\dagger$ and this yields
\eqnlb{
\log\det{i\di\ov i\fdi}=-{1\ov 4\pi}\int \Big((\nabla\varphi,\nabla\varphi)
+\al\bar\al\Big).}{baby2}
The $3$-dimensional integral in \refs{def16} converted into an
ordinary $2$-dimensional spacetime integral because we have chosen
a triangular $h$ in the decomposition \refs{baby1}.
The property that the Wess-Zumino term becomes local for a
triangular $h$ has been exploited in a different context in \cite{WZNW}.\par
At this point we wish to comment on the $\theta$-independence
of the fermionic determinant. For $v\es 0$ this fact is
easily understood as follows:\pan
In this case $A_\mu=\ha\eps_{\mu\nu}\pa_\nu\varphi\sigma_3$ and
$i\di$ is just the tensor product of two $U(1)$ Dirac operators,
one with $\varphi\to\ha\varphi$ and the other with
$\varphi\to -\ha\varphi$. This means that the $\log\det$
is just the sum of the two abelian results
with the corresponding replacements and in this sum
the $\theta$-dependent terms cancel.
In section 6 we have shown that this cancellation between the
various colour degrees of freedom takes actually place
for arbitrary gauge potentials and semi-simple gauge
groups.\par
With the parametrization \refs{baby1} there is actually a much quicker
way to arrive at \refs{baby2}. When we replace
$\varphi,v$ in \refs{baby1} and in
\eqnn{
F_{01}=-\ha\pmatrix{\lap\varphi+\al\bar\al&\bar\pa\al-\al\bar\pa\varphi\cr
\pa\bar\al-\bar\al\pa\varphi&-\lap\varphi-\al\bar\al\cr},\mtxt{
where} \al=\pa v+v\pa\varphi}
by the deformed fields $\tau\varphi,\tau v$, then $F_{01}$ and
\eqnn{
a+a^\dagger=-\pmatrix{\varphi & v(1+\tau\varphi)\cr
\bar v(1+\tau \varphi)&-\varphi\cr}}
in \refs{def15} both become polynomial in $\tau$
and the $\tau$-integral can easily be performed.\par
Similar as the fermionic determinant the Yang-Mills action
\eqnl{S_{YM}={1\ov 2 g^2}\int\limits_\cm \tr F_{01}^2=
{1\ov 4g^2} \int\Big\{(\pa\bar\pa\varphi+\al\bar\al)^2+
\vert\bar\pa\al-\al\bar\pa\varphi\vert^2\Big\}}{baby3}
depends on $v$ only via the $\alpha$-field and this suggests that
we should change variables $A^a_\mu\to (\varphi,\al,\bar\al,U)$.
To find the Jacobian of this transformation we note that,
up to a gauge transformation,
\eqnn{
A_z=i\pmatrix{\ha\pa\varphi&\al\cr0&-\ha\pa\varphi\cr}+i\pa UU^{-1}}
and parametrize the gauge transformations as
\eqnn{
U=U(\xi)\Rightarrow \pa_\mu U U^{-1}={\cal N}_{ab}\tau_a\pa_\mu\xi^b
,\mtxt{where}{\cal N}_{ab}=2\tr\Big({\pa U\ov\pa\xi^b}U^{-1}\tau_a\Big)}
and the $\tau_a$ are half of the Pauli-Matrices. Then the transformation
to the new variables is given by
\eqnn{
\pmatrix{A^a_0\cr A^a_1\cr}=
\pmatrix{0&0&-1&&&\cr 0&-1&0&&{\cal N}\pa_0&\cr \pa_1&0&0&&&\cr
0&-1&0&&&\cr 0&0&1&&{\cal N}\pa_1&\cr
-\pa_0&0&0&&&\cr}\pmatrix{\varphi\cr\al\cr\bar\al\cr\xi^1\cr\xi^2\cr\xi^3\cr}}
and we conclude that the Jacobian of this transformation depends
only on $U$,
\eqnl{
\cd A=J(U)\cd\varphi\cd\al\cd\bar\al\cd U\mtxt{,}
J(U)\cd U\sim \det\lap d\mu(U).}{baby3a}
When calculating expectation values of gauge
invariant operators the factor $\det\lap$ und the integrations over
the Haar measure $d\mu(U)$ in the numerator and denominator cancel.\par
In particular for the chiral condensate \refs{corr3}
in $N_f$-flavour baby-$QCD$ we find
\eqnl{
\<\bar uP_+u\>(x)=-S^\theta_{++}(x,x;0){\int \cd(\varphi,\al,\bar\al)
\,\tr J\,e^{-S_{YM}}{\det}^{N_f}(i\di)\ov
\int \cd(\varphi,\al,\bar\al)\,e^{-S_{YM}}{\det}^{N_f}(i\di)}}{baby4}
or after inserting the explicit expressions we are left with
the non-Gaussian functional integral
\eqnl{
\<\bar uP_+ u\>=-{e^\theta\ov 2\pi R}{1\ov 1-r^2/R^2}
{\int \cd (.)\,\Big\{e^\varphi (1+v\bar v)+e^{-\varphi}\Big\}\,e^{-\Gamma}
\ov \int \cd (.)\, e^{-\Gamma},}}{baby5}
with effective action
\eqnl{
\Gamma=S_{YM}+{N_f\ov 4\pi}\int\Big\{(\nabla\varphi,\nabla\varphi)+\al\bar\al
\Big\}.}{baby6}
Thus we have reduced the task of calculating the 'quark' condensate
to computing the functional integral \refs{baby5} over the
gauge invariant variables $\varphi$ and $\al$. For an evaluation
of the integral it maybe relevant to decide on the boundary
conditions for the gauge fields. For the abelian models it
makes no difference whether we take free boundary conditions
or impose the gauge invariant bag boundary conditions \cite{bc1}
\eqnn{n^\mu F_{\mu\nu}\vert_{\cmb}=0,}
but for the non-abelian theories this choice may affect the final
results for correlators.\par
The formula \refs{baby6}
immediately leads to a gauge invariant perturbation expansion
for the condensate and similarly for other expectation values.
Note that if we perturb about the quadratic part of the
effective action then we obtain an infinite resummation of the ordinary
perturbative expansion in the gauge coupling constant.
We hope to report on the corresponding results elsewhere.
Here we shall truncate the nonabelian theories and shall
investigate their abelian projections.
\subsection{Abelian projection of $SU(N_c)$ gauge theories.}
Here we calculate the condensate in the approximation where
the 'gluons' are confined to the Cartan subalgebra of $SU(N_c)$.
Hence only $N_c\!-\!1$ gluons propagate around a 'gluon' loop
and there are no $3$ or $4$-gluon vertices in this approximation.
In other words, we assume that $ g$ in $ A_z=i g^{-1}\pa_z
 g$ lies in the maximal abelian subgroup of $SL(N_c)$, i.e.
\eqnl{
 g=\prod_{i=1}^{N_c-1} e^{-g(\varphi_i+i\lam_i)
H_i}}{nagt5}
with trace-orthonormal $H_i$ in the Cartan subalgebra of $SU(N_c)$.
The Jacobian of the transformation $( A_\mu)
\to (\varphi_i,\lam_i)$, where $ A$
lies in the Cartan subalgebra, is field independent
and cancels in expectation values of gauge invariant
observables. Thus in the abelian projected theory the 'quark'
condensate \refs{corr3} simplifies to
\eqnl{
\<\bar uP_+u\>_{SU(N_c)}=-{e^\theta\ov 2\pi R}{1\ov 1-r^2/R^2}
\;\tr\prod_{i=1}^{N_c-1}{\int\cd\varphi_i\,e^{-2g\varphi_iH_i}
\,e^{-\Gamma_0[\varphi_i]}\ov
\int\cd\varphi_i \,e^{-\Gamma_0[\varphi_i]}},}{nagt6}
where $\Gamma_0$ is the effective action $\Gamma_\theta$ in
\refs{def21} without boundary term ($\theta\es 0$), with $e$ replaced
by $g$ and with $N_c\es 1$. The $N_c\!-\!1$ functional integrals
can be calculated by using that
\eqnn{
{\int \cd\varphi\, e^{-2g\varphi H}e^{-\Gamma_0}\ov
\int \cd\varphi\,e^{-\Gamma_0}}=
\Big({\tilde mRe^{\gam+F_g}\ov 2}[1-{r^2\ov R^2}]\Big)^{H^2/N_f},}
where now $\tilde m^2\es N_fg^2/\pi$ and $F_g$ is the function \refs{lll7c}
with the electric charge $e$ replaced by the gauge coupling $g$ or
equivalently $m_\eta$ by $\tilde m$. Since $N_c\sum H_i^2\es
(N_c\!-\!1)I_c$ we arrive at the following expression for the chiral
condensate in the projected theories
\eqnl{
\<\bar uP_+u\>_{SU(N_c)}=-{e^\theta\ov 2\pi R}{N_c\ov 1-r^2/R^2}\Big(
{\tilde m Re^{\gam+F_g}\ov 2}[1-{r^2\ov R^2}]\Big)^{(N_c-1)/N_cN_f}.}{nagt7}
In the one-flavour model the condensate depends on the bag-radius
as $\sim R^{-1/N_c}$ and therefore saturates the upper bound \refs{nagt4a}.\par
The $U(N_c)$-condensate is related to the one in $SU(N_c)$ gauge
theories as in \refs{nagt3} and thus is found to be
\eqngrl{
\<\bar uP_+u\>_{U(N_c)}&=&-{e^{\theta I_e}\ov 2\pi R}{N_c\ov 1-r^2/R^2}
\big({e\ov g}\Big)^{1/N_cN_f}\,e^{(F_e-F_g)/N_cN_f}}
{&&\cdot\Big({\tilde m Re^{\gam+F_g}\ov 2}[1-{r^2\ov R^2}]\Big)^{1/N_f}.}
{nagt8}
Let us now discuss the various limiting cases in turn.
\paragraph{Large $N_c$ limit.}
The large $N_c$ limits of the ablian projected theories are
different from the same limits in the full theories since there
is no suppression of fermionic loops relative to the bosonic
ones. But as in the full theories a condensate remains
in the thermodynamic limit in the one-flavour models.
Indeed, when $N_c\to\infty$ the condensates
at the center of a large bag simplify to
\eqnl{
\<\bar uP_+u\>_{SU(N_c)}=e^\theta \<\bar uP_+ u\>_{U(N_c)}
=-{e^\theta N_c\ov 2\pi R}\Big({\tilde m R e^\gam\ov 2}\Big)^{1/N_f}.}{nagt9}
For $N_f\es 1$ a condensate remains for infinite volume
and its limiting value is just
\eqnl{
{1\ov N_c}\<\bar uP_+u\>_{SU(N_c)}={e^\theta\ov N_c}
\<\bar uP_+ u\>_{U(N_c)}=-{e^{\theta+\gam} g\ov 4\pi^{3/2}}.}{nagt10}
\paragraph{Weak couplings.}
For a small electric charge $e$ the function $I_e$ in the
first factor in \refs{nagt8} tends to $1$ and inserting
the asymptotic expansion \refs{lll10} for small $m_\gam R$
we see that for $e\to 0$ the $U(N_c)$-condensate converges
to the $SU(N_c)$ one, as expected.\pan
When the gauge coupling $g$ is weak the $SU(N_c)$-condensate
becomes equal to $-N_c$ times the chirality violating entry $S^\theta_{++}$
of the free Green's function \refs{green4} and thus vanishes
in the thermodynamic limit. The $U(N_c)$-condensate
simplifies to $N_c$ times the $U(1)$ condensate \refs{nagt2}.
\paragraph{Strong couplings.}
When both couplings $e$ and $g$ become strong, or equivalently
the bag very large, then the condensates at the bag center
are just
\eqngrl{
\<\bar uP_+u\>_{SU(N_c)}&=&-{e^\theta N_c\ov 2\pi R}\Big(
{\tilde m Re^\gam\ov 2}\Big)^{(N_c-1)/N_cN_f}}
{\<\bar uP_+u\>_{U(N_c)}&=&-{N_c\ov 2\pi R}
\big({e\ov g}\big)^{1/N_cN_f}\Big(
{\tilde m Re^\gam\ov 2}\Big)^{1/N_f}.}{nagt11}
\section{Discussion}
In this paper we have investigated Euclidean gauge theories with
massless Dirac fermions enclosed in a bag. We have imposed
$U_A(N_f)$-breaking boundary conditions to trigger a
breaking of the chiral symmetry. In the first part
of the paper we considered gauge theories in arbitrary
$2n$-dimensional bags. We found the explicit $\theta$-dependence
of the fermionic Green's functions and determinants in
arbitrary background gauge fields. In contrast to the
situation on a sphere or torus the Dirac operator possesses no zero
modes in a bag and this property simplifies the quantization
considerably. In the second part of the paper we investigated
$2$-dimensional gauge theories. We found the mesonic current
correlators and calculated the chiral condensates both
for abelian and non-abelian gauge theories. Our results
are in full agreement with earlier instanton-type or
small 'quark'-mass calculations. We conclude that the
bag boundary conditions are a substitute for introducing
small quark masses to drive the breaking of the chiral
symmetry. Of course, for several flavours the condensate
dissappears when the volume of the bag tends to infinity,
in accordance with general theorems. Only when the number
of colours is sent to infinity before the thermodynamic
limit is performed there remains a 'quark'-condensate.\par
On a sphere or torus one finds that in the chiral limit
only configurations with vanishing topological charge
\eqnn{
q={e\ov 2\pi}\int d^2x\,F_{01}\mtxt{resp.}
q={g^2\ov 32\pi^2}\int d^4x\,\eps_{\mu\nu\al\beta}\tr F_{\mu\nu}
F_{\al\beta}}
contribute to the partition functions in $2$ resp. $4$-dimensions
\cite{LeutSm}. For $U(N_c)$ gauge theories confined in a
$2$-dimensional bag we can find the expectation values of
arbitrary powers of the topological charge by differentiating the
partition function sufficiently often with respect to $\theta$.
The correlators are reproduced by the following
Gaussian distribution for the topological charge:
\eqnl{
d\mu(q)=\sqrt{N_cN_f\ov \pi \sigma}\,e^{-N_cN_f \sigma[q+\theta/2\sigma]^2}
\,dq\mtxt{,}\sigma={I_0(m_\eta R)\ov m_\eta R I_1(m_\eta R)}.}{d1}
The expectation value of the instanton number vanishes
for vanishing $\theta$, but its fluctuation does not.
Only for very small volumes and/or weak coupling (for which
the semiclassical approximation makes sense)
is the instanton number distribution sharply peaked
about $q\es 0$ as can be seen by inspection from
\refs{d1} or from
\eqnl{
\<\vert q\vert\>=\left\{ \begin{array}{ll}
0&\hbox{for }m_\eta R\to 0\\
\sqrt{eR\ov \pi N_c}(\pi N_f)^{-1/4}&\hbox{for }
m_\eta R\to\infty.\end{array} \right.   }{d2}
For big volumes and/or strong coupling, which would correspond to small quark
masses, configurations with $q^2\sim 1/\sqrt{N_f}$
dominate the functional integral.\par
In this paper we have regarded the bags as mathematical
constructs rather than real objects in spacetime. For example,
to be a model for a hadron at finite temperature, $\cm$ must
be a bag in space and hence $[0,\beta]\times \cm$ a subspace
of the Euclidean spacetime. The gluon (quark) fields must then
be periodic (antiperiodic) in the Euclidean time with period
$\beta\es 1/T$. In \cite{DW} we have studied multi-flavour $QED_2$
at finite temperature enclosed in a spatial bag $[0,L]$.
Besides the finite temperature boundary conditions we imposed
the bag boundary conditions $B_\theta\psi\es \psi$ at
$x^1\es 0$ and $x^1\es L$. By applying the methods developped in this paper
we found for the chiral condensate in the low temperature
limit $T\ll 1/L\ll m_\eta$ \cite{DW}
\eqnl{
\<\bar u P_+u\>=-{1\ov 4L}\, e^{\gam/N_f}\Big({m_\eta L\ov \pi}\Big)^{1/N_f}
}{d3}
In particular, for $2$ flavours this reads
\eqnl{
\<\bar u P_+u\>=-\Big({e^\gam m_\eta \ov 16 \pi L}\Big)^{1/2}}{d4}
and this result is identical to that of Shifman and
Smilga \cite{ShifSmil} when they allowed for fracton
configurations.\par
The condensate in an $d$-dimensional Euclidean bag obeys the scaling relation
\cite{WW}
\eqnl{
\<\bar\psi P_+\psi\>\big(\lam R,\lam x,g\big)=\lam^{1-d}\,Z(\lam)\<\bar\psi
P_+\psi\>\big(R,x,\lam^{2-d/2}g(\lam)\big),}{d5}
where $Z(\lam)$ and $g(\lam)$ are the wave-function renormalization
of the condensate and running gauge coupling constant, respectively\footnote{
up to possible runnings of the surface coupling constants}. The relative
size $\lam$ of the two bags plays the role of the inverse
energy scale in the Callan-Symanzik equation. For example,
the condensates in the multi-flavour Schwinger models, \refs{lll9},
obey this scaling relation with $g(\lam)\es g$ and $Z(\lambda)\es 1$
and this agrees with the wellknown fact that the
$\beta$-function vanishes and that there is no wave function
renormalization in these theories. In $4$-dimensions $g(\lam)$
becomes weak in small bags because of asymptotic freedom
and the chiral condensate should again tend to the chirality
violating entry $S_{++}^\theta$ of the free Green's function.
The change of the condensate at $x\es 0$, when the size of the bag is
increased, is then determined by the nonperturbative beta-function
and anomalous dimension of the condensate. Thus we could extrapolate
the $QCD$-condensate to large volumes if we would know its
anomalous dimension and the $QCD$ beta-function. Conversely,
we may put bounds on the functions $g(\lambda), Z(\lambda)$
since a condensate must remain in the infinite volume limit.\par
\paragraph{Acknowledgements:} The reported work was partially
supported by the Swiss National Science Foundation. We are
indebted to A. Abrikosov jr., S.I. Azakov, I. Sachs and C. Wiesendanger
for valuable discussions and to J. Fr\"ohlich for pointing our
attention to Ref. \cite{spin}.
\begin{appendix}
\section{Appendix}
In this appendix we fill the gaps in the calculation of
the fermionic determinants confined in $2$-dimensional bags
in section $6$. What remains is to calculate the surface
Seeley deWitt coefficent $b_1$ in \refs{def14} which enters
in (\ref{det6},\ref{def12}).\par
First we note that $\oint \tr b_n(\phi)$ has the expansion
\eqnl{
\oint \tr b_n(\phi)=\sum\limits_0^{d-1}\,\oint \tr c_p(F_{\mu\nu},{\cal
R},\chi)
\,\pa_n^p\phi,}{a1}
where $c_p$ is a gauge- and Lorentz-invariant
local polynomial in the field-strength and its covariant derivatives,
the extrinsic and intrinsic curvatures of the bag boundary and their
covariant derivatives and has length-dimension $1-d+p$.
Here $\pa_n^p$ is the $p$'th derivative normal to the bag
boundary. In particular in two dimensions we need $b_1$
which is the sum of two terms (again neglecting purely
geometric contributions)
\eqnl{
\oint \tr b_1(\phi)=\oint \tr f_1(\theta)\chi\,\phi+
\oint \tr f_2(\theta)\,\pa_n\phi.}{a2}
Here we are not interested in the term containing $f_1$. In \refs{def12}
it would not contribute since $A\!+\!A^\dagger$ vanishes on the
bag boundary and in \refs{det6} it would yield an uninteresting
constant which cancels in expectation values\footnote{
It would contribute to the free energy or to the
Casimir effect \cite{Falo}.}
The invariance of the fermionic determinant under parity,
$(\theta,A,x)\to (-\theta,\tilde A,\tilde x)$, restricts
the form of the free function $f_2$.
To determine this function it suffices to calculate the heat kernel
expansion for free fermions confined to the halfplane
$\cm =\{x^0,x^1\vert x^1\geq 0\}$ and subject to bag boundary conditions
at $x^1\es 0$.\par
Besided the wellknown properties the heat kernel must obey
the boundary conditions
\eqngr{
B_\theta K(t,x,y)\vert_{x^1=0}&=&K(t,x,y)\vert_{x^1=0}}
{B_\theta \fdi_x K(t,x,y)\vert_{x^1=0}&=&\fdi_x K(t,x,y)\vert_{x^1=0}.}
After some algebra we have found the following explicit formula
\eqngrr{
&&K(t,x,y)={1\ov 4\pi t}e^{-(\xi_0^2+\xi_1^2)/4t}}
{&&\quad+{1\ov 4\pi t}\pmatrix{e^\theta\sinh\theta&-\cosh\theta\cr
-\cosh\theta&-e^{-\theta}\sinh\theta\cr}\;
e^{-(\xi_0^2+\eta^2)/4t} }
{&&\quad+{i{\cal P}\sinh\theta\ov 8t\sqrt{\pi t}}
\pmatrix{e^\theta& -1\cr -1&e^{-\theta}\cr}
e^{-{\cal P}^2/4t}\Bigg[1+\hbox{erf}\Big({i\xi_0\sinh\theta-\eta\cosh\theta\ov
2\sqrt{t}}\Big)\Bigg],}
where $\xi^\mu\es x^\mu-y^\mu,\;\eta\es x^1+y^1$ and
${\cal P}=\xi_0\cosh\theta+i\eta\sinh\theta$.
To determine the relevant Seeley-deWitt coefficient we need
to calculate
\eqnl{
\int\limits_\cm K(t,x,x)f(x)
\sim\!\int\limits_\cm K(t,x,x)\Big(f(x^0,0)-x^1\pa_1 f(x^0,0)
+..\Big),}{a4}
where we anticipated that the integrand is sharply peaked
at $x_1\es 0$ and thus expanded the test function $f$ about
$x^1\es 0$.
On the diagonal ($x\es y$) we have $\xi\es 0$ and $\eta\es 2x_1$ and
we are left with calculating the integrals
\eqngr{
&&\int\limits_{x_1\geq 0}dx_1\, e^{-x_1^2/t}\Big(f(x_0,0)+x_1\pa_1
f(x_0,0)+..\Big)}
{&&\int\limits_{x_1\geq 0}dx_1\,x_1\,e^{x_1^2\sinh^2\theta/t}\Big[
1-\hbox{erf}\big({x_1\cosh\theta\ov \sqrt{t}}\big)\Big]\Big(
f(x_0,0)+x_1\pa_1 f(x_0,0)+..\big).}
The first integral is easily evaluated by using that
\eqnn{
\int\limits_{x\geq 0}dx\, e^{-x^2/t}=\ha \sqrt{\pi t}\quad,\quad
\int\limits_{x\geq 0} dx\,x e^{-x^2/t}={t\ov 2}.}
For evaluating the second integral we need the formulae
\eqngr{
\int\limits_0^\infty dx\,
\big[1-\hbox{erf}(\beta x)\big]\,e^{\mu x^2}\,x
&=&-{1\ov 2\mu}\Big(1-{\beta\ov\sqrt{\beta^2-\mu}}\Big)}
{\int\limits_0^\infty dx\,
\big[1-\hbox{erf}(\beta x)\big]\,e^{\mu x^2}\,x^2
&=&{1\ov 2\mu\sqrt{\pi}}\Big({\beta\ov \beta^2-\mu}+
{1\ov 2\sqrt{\mu}}\log{\beta-\sqrt{\mu}\ov \beta+\sqrt{\mu}}\Big)}
which apply if $\mu>0$ and $\Re (\mu)<\Re (\beta^2)$.
Using these results one finds the following small-$t$ expansion
for the integral \refs{a4}
\eqngrr{
&&\int d^2x K(t,x,x)f(x)={1\ov 4\pi t}\int\limits_\cm d^2x f(x)}
{&&\quad+{1\ov 8\sqrt{\pi t}}\int dx^0\Bigg\{\pmatrix{
e^\theta&-1\cr -1&e^{-\theta}\cr}-I\Bigg\}\,f(x^0,0)}
{&&\quad+{1\ov 8\pi} \int dx^0 \Bigg\{{\log e^\theta\ov \sinh\theta}
\pmatrix{e^\theta&-1\cr -1&e^{-\theta}\cr}-I\Bigg\}\,\pa_1 f(x^0,0)
+O(t^{1/2}).}
The first term on the right yields the wellknown $a_0$ coefficient,
the second term $b_{1/2}$ and the third one is the
$b_1$-coefficient \refs{def14} (after noting that
$\pa_1\es -\pa_n$) we have been aiming at.
We see that the small $t$-expansion of $K$ is invariant under $\theta\to
\theta+i2\pi n,\; n\in Z$, as required.
\end{appendix}

\end{document}